\documentclass[final,3p,times]{elsarticle}
\usepackage{graphicx}
\usepackage{amssymb}
\usepackage{amsfonts}
\usepackage{amsmath}
\usepackage{color}
\usepackage{amsmath}
\usepackage{float}
\usepackage{mathtools}
\usepackage{algorithm}
\usepackage{algpseudocode}

\usepackage[colorinlistoftodos]{todonotes}
\usepackage[colorlinks=true, allcolors=blue]{hyperref}
\usepackage{subcaption}
\usepackage{mwe}
\usepackage{epstopdf}

\newcommand{\e}{\varepsilon}
\newcommand{\s}{\sigma}

\newcommand{\qi}{q^j_l}
\newcommand{\qj}{q^j_m}
\newcommand{\qv}{\mathbf{q}^n}
\newcommand{\ra}{A_{{\rm R},j}}
\newcommand{\ia}{A_{{\rm I},j}}
\newcommand{\rx}{B_{{\rm R},j}}
\newcommand{\ix}{B_{{\rm I},j}}
\newcommand{\curlB}{[\nabla \times \mathbf{B}(\mathbf{r})]_z}

\setcitestyle{authoryear,round}

\begin{document}

\begin{frontmatter}

\title{A coarse-grained phase-field crystal model of plastic motion}

\author[TUD,HKIAS]{Marco~Salvalaglio\corref{corrauth}}
\cortext[corrauth]{Corresponding author}
\ead{marco.salvalaglio@tu-dresden.de}
\author[OSLO]{Luiza~Angheluta}
\author[WAY]{Zhi-Feng~Huang}
\author[TUD,DCMS]{Axel~Voigt}
\author[OAK]{Ken~R.~Elder}
\author[MIN]{Jorge~Vi\~nals}

\address[TUD]{Institute of Scientific Computing, Technische Universit\"at Dresden, 01062 Dresden, Germany.}
\address[HKIAS]{Hong Kong Institute for Advanced Studies and Department of Materials Science and Engineering, City University of Hong Kong, Hong Kong SAR, China.}
\address[OSLO]{PoreLab, The Njord Centre, Department of Physics, University of Oslo, P. O. Box 1048, 0316 Oslo, Norway}
\address[WAY]{Department of Physics and Astronomy, Wayne State University, Detroit, Michigan 48201, USA}
\address[DCMS]{Dresden Center for Computational Materials Science (DCMS), Technische Universit\"at Dresden, 01062 Dresden, Germany}
\address[OAK]{Department of Physics, Oakland University, Rochester, Michigan 48309, USA}
\address[MIN]{School of Physics and Astronomy, University of Minnesota, 116 Church Street SE, Minneapolis, Minnesota 55455, USA}

\begin{abstract}
The phase-field crystal model in an amplitude equation approximation is shown to provide an accurate description of the deformation field in defected crystalline structures, as well as of dislocation motion. We analyze in detail stress regularization at a dislocation core given by the model, and show how the Burgers vector density can be directly computed from the topological singularities of the phase-field amplitudes. Distortions arising from these amplitudes are then supplemented with non-singular displacements to enforce mechanical equilibrium. This allows for a consistent separation of plastic and elastic time scales in this framework. A finite element method is introduced to solve the combined amplitude and elasticity equations, which is applied to a few prototypical configurations in two spatial dimensions for a crystal of triangular lattice symmetry: i) the stress field induced by an edge dislocation with an analysis of how the amplitude equation regularizes stresses near the dislocation core, ii) the motion of a dislocation dipole as a result of its internal interaction, and iii) the shrinkage of a rotated grain. We compare our results with those given by other extensions of classical elasticity theory, such as strain-gradient elasticity and methods based on the smoothing of Burgers vector densities near defect cores.
\end{abstract}


\end{frontmatter}

\section{Introduction}
The distribution of dislocations and grain boundaries and their motion play an important role in materials physics as they determine many material properties and response, especially in polycrystalline and heteroepitaxial systems. A great deal of research has been devoted to the study of such systems to better understand the interplay between the many disparate length scales involved \citep{Rollett2015,Sethna2017}. Microscopic theories, such as Density Functional Theory and Molecular Dynamics, provide detailed descriptions at the microscopic scale, but are unfortunately restricted to relatively small length and time scales. Coarse-grained methods have also been introduced, such as Discrete Dislocation Dynamics (DDD) \citep{Kubin92,Devincre1992}, which evolve dislocation lines through Peach-Koehler type forces. These methods can examine mechanical properties on large length and time scales while treating dislocations explicitly. However, they rely on the phenomenology involved in dislocation line mobilities and their reconnections. Phase-field (PF) models also belong to the class of coarse-grained approaches. They are based on the description of distinct phases via continuous order parameters, and the implicit description of interfaces between them. Early work on the application of PF models to the description of extended defects and, in turn, to their motion, focused on describing elementary defects as an eigenstrain, which is then mapped onto a set of phase fields \citep{re:wang01,re:koslowski02,RODNEY2003,re:bulatov06,re:wang10}. As discussed extensively in \cite{re:wang10}, a fundamental advantage of this methodology is that physical parameters describing dislocation energies are subsumed in the energy model, and hence do not need separate specification (e.g, generalized stacking fault energies). The evolution of the phase fields is purely dissipative, and is driven by minimization of a phenomenological energy of the Ginzburg-Landau type. However, a few quantities that remain central to the analysis of defect motion still require a priori specification, such as active slip systems and defect mobilities. As we will argue below, the phase field crystal (PFC) model overcomes these limitations.

The PFC model \citep{Elder2002,Elder2004,Emmerich2012} is based on the definition of a continuous order parameter or field $\psi$ that is related to the atomic number density. A free energy functional is introduced that is minimized by a periodic field with the desired lattice symmetry. The temporal evolution of $\psi$ is assumed to be governed by conserved and dissipative dynamics, hence changing on effectively diffusive time scales. The formulation is akin to Classical Density Functional Theory, and it has been justified to some extent on those grounds as well \citep{Elder2007,vanTeeffelen2009,re:archer19}. Therefore this modeling is intermediate between fully atomistic and classical continuum theories. Diffusive time scales and atomistic length scales allow for describing grain boundaries in detail, their motion, and the response of complex polycrystalline configurations, as these phenomena are well described by energy relaxation alone, without requiring the calculation of any stress fields in the grains \citep{re:li18,re:wu09,backofen14,re:mianroodi15,re:hirvonen16,re:clayton16,koehler16}. By focusing on the complex slowly-varying amplitudes of $\psi$, a spatial coarse-grained version of the PFC model has been introduced \citep{Goldenfeld2005,Athreya2006,GoldenfeldJSP2006,Yeon2010}. This amplitude expansion (APFC) approach provides an approximate description of the atomic length scale \citep{Spatschek2010,ElderPRE2010,Heinonen2014,Salvalaglio2019}. Due to its coarse grained nature and the use of adaptive numerical methods, computational costs can be drastically reduced and systems that are orders of magnitude larger than those accessible with PFC models can be described with APFC models \citep{Praetorius2019}. A different methodology but analogous in many respects is the Field Theory of Dislocation motion \citep{re:acharya01,re:zhang15,re:zhang18b}, and its more recent nanoscale implementation in the Generalized Disclination Theory \citep{re:zhang18}.

At variance with models explicitly tracking defects and grain boundaries, the (A)PFC model starts from the specification of the governing free energy from which lattice symmetry, possible defects, and their combination rules, follow. Despite these advantages, a long-standing difficulty associated with the (A)PFC model is that the evolution of the field is diffusive, hence elastic response in the time scale appropriate for plastic motion is incomplete. The first attempt to overcome this problem was the development of the so-called modified PFC (MPFC) model given in \citep{Stefanovic06}, in which a higher-order time derivative was introduced into the equation of motion.  While this approach does lead to faster elastic relaxation, it gives incorrect behavior in the large wavelength limit, as pointed out in \citet{Majaniemi07} and \citet{Heinonen2016}. More recently, a complete transport theory has been developed in which the order parameter that enters the PFC model is included as a constitutive component within the more general laws of balance of mass and momentum \citep{Heinonen2016}. While these models do introduce a \lq\lq fast" time scale on the order of the speed of sound, they can become computationally expensive when the time scale of mechanical relaxation is orders of magnitude faster than the time scales for mass diffusion and plastic distortion. An alternative and computationally convenient method seeks to enforce elastic equilibrium through an interpolation scheme in the PFC model designed to achieve fast mechanical relaxation \citep{Zhou19}. The method, however, is limited to uniaxial external deformation. Yet another approach in the APFC model \citep{Heinonen2014} recognizes that the phases of the complex amplitudes contain information about the elastic distortion and that instantaneous mechanical equilibrium can be achieved by relaxing these fields at a faster rate.  Similar to other approaches, this limits the computational efficiency of the method and lacks transparency. A more recent approach that retains the atomic density as constitutively governing plastic slip but adds the elastic distortion caused by lattice incompatibility was described in  \cite{Skaugen2018b,Skaugen2018}. This latter approach is also computationally efficient, and provides for the connection between the phase field and elasticity theory in the presence of dislocations. We will extend this latter approach to the APFC model in order to provide a computational approach that is suitable for larger systems and spatial scales.

In this work, we thus present a coarse-grained description of deformations in crystals and plasticity based on the APFC model, intermediate between atomistic and continuum length scales. Section \ref{sec:apfc} summarizes the equations governing the evolution of the PFC model and the associated description based on the slowly varying complex amplitudes, i.e., on the APFC model. While the expansion in a slowly varying amplitude assumes prior knowledge of the lattice symmetry, the latter simply follows from the minimization of the free energy functional defining the system of interest. This description is advantageous from a computational point of view as it does not need to resolve the variation of the order parameter at the scale of the underlying lattice parameter as in the PFC model. It is limited, however, to small distortions away from a reference lattice, not an uncommon restriction in this class of studies, as it focuses on length scales larger than the atomic spacing. Section \ref{sec:corr} addresses how, in the APFC model, stress fields in the presence of defects are calculated, and how mechanical equilibrium is enforced on the time scale of defect motion. A finite element implementation of the combined system of equations is provided in Sec. \ref{sec:apfc_fem}, followed by our numerical results in Sec. \ref{sec:results}. We begin by studying the stress field created by an isolated, stationary dislocation in a two-dimensional, large crystal (Sec. \ref{sec:single}). The computed stress agrees with classical elasticity far from the defect core, and also with results provided by either first strain gradient elasticity \citep{Lazar2005}, or Burgers vector smoothing methods \citep{Cai2006} near the core. Sections \ref{sec:dipole} and \ref{sec:dipole-motion} present our results on dislocation dipoles and their motion under each other's influence. Sec. \ref{sec:grain-shrinkage} addresses the study of grain shrinkage along with the analysis of defect distribution achieved within the presented framework.

\section{Amplitude Phase-Field Crystal (APFC) model}
\label{sec:apfc}

The PFC model describes a crystal lattice by means of a continuous, periodic field $\psi$, the dimensionless atomic probability density difference \citep{Elder2002,Elder2004,Emmerich2012}. A phenomenological free energy is introduced as
\begin{equation}
F_\psi=\int_{\Omega} \left[\frac{\Delta B_0}{2}\psi^2+\frac{B^x_0}{2} \psi(1+\nabla^2)^2\psi
-\frac{t}{3}\psi^3+\frac{v}{4}\psi^4 \right]d\mathbf{r},
\label{eq:F_PFC}
\end{equation}
which describes a first order transition between a disordered/liquid phase, where $\psi$ is uniform, and an ordered/crystalline phase, where $\psi$ is periodic in the domain $\Omega$. $\Delta B_0$, $B_0^x$, $v$ and $t$ are parameters determining which phase minimizes the free energy $F_\psi$ \citep{Elder2007}. In the standard approach, the evolution towards equilibrium for out-of-equilibrium configurations is described by the gradient flow ensuring conservation of $\psi$
\begin{equation}
\frac{\partial \psi}{\partial t}= M \nabla^2 \frac{\delta F_\psi}{\delta \psi},
\label{psitime}
\end{equation}
where $M$ is a mobility. In the crystalline state, $\psi$ can be generally approximated as a sum of plane waves, i.e.,
\begin{equation} 
    \psi = \psi_0+\sum_\mathbf{q} A_\mathbf{q} e^{i\mathbf{q}\cdot \mathbf{r}}=\psi_0+\sum_j^N A_j e^{i\mathbf{q}_j\cdot \mathbf{r}} + {\rm c.c.},
    \label{eq:psi}
\end{equation}
where $\psi_0$ is the average density, set to zero in the following, $A_j$ are the (complex) fields corresponding to the amplitudes of each plane wave and $\{ \mathbf{q}_j \}$ is the set of $N$ reciprocal lattice vectors representing a specific crystal symmetry.

In the so-called amplitude expansion of the PFC model (APFC) \citep{Goldenfeld2005,Athreya2006,GoldenfeldJSP2006}, the crystal structure is described directly by means of $A_j$. They account for distortions and rotations of the crystal structure with respect to a reference state accounted for by a proper set of $\mathbf{q}_j$ vectors. Under the assumption of slowly varying amplitudes, the free energy of the system expressed in terms of $A_j$ reads  
\begin{equation}
F_A=\int_{\Omega} \bigg[\frac{\Delta B_0}{2}\Phi+\frac{3v}{4}\Phi^2 +\sum_{j=1}^N
\left ( B_0^x |\mathcal{G}_j A_j|^2-\frac{3v}{2}|A_j|^4 \right ) 
+f^{\rm s}(\{A_j\},\{A^*_j\}) \bigg]  d \mathbf{r}, 
\label{eq:energyamplitude}
\end{equation}
where $\mathcal{G}_j = \nabla^2+2i\mathbf{q}_j \cdot \nabla$ and $\Phi = 2\sum_{j=1}^N |A_j|^2$. The term $f^{\rm s}(\{A_j\},\{A_j^*\})$ corresponds to a complex polynomial of $A_j$ and $A_j^*$ and is determined by the crystalline symmetry of the reference lattice \citep{ElderPRE2010,SalvalaglioPRE2017}. Amplitude functions allow for a full description of elastic deformation within the PFC framework, and the associated energy in Eq. \eqref{eq:energyamplitude} contains the elastic energy associated with deformations \citep{ElderPRE2010,Heinonen2014}. The evolution equation of the $A_j$'s can be derived from Eq.~\eqref{psitime}, and reads
\begin{equation}
\frac{\partial A_j}{\partial t} =-|\mathbf{q}_j|^2 \frac{\delta F_A}{\delta A_j^*},
\label{eq:amplitudetime}
\end{equation}
with time rescaled by $M$. In this work we focus on two-dimensional crystals with triangular symmetry ($N=3$) described by
\begin{equation}
\begin{gathered}
\mathbf{q}_1=k_0 \left(-\sqrt{3}/2,-1/2 \right),\ \
\mathbf{q}_2=k_0(0,1), \ \
\mathbf{q}_3=k_0\left(\sqrt{3}/2,-1/2 \right), 
\end{gathered}
\label{eq:ktri}
\end{equation}
with $k_0=1$, while $f^{\rm s}(\{A_j\},\{A_j^*\})=-2t(A_1A_2A_3+A_1^*A_2^*A_3^*)$.

\section{APFC dynamics and mechanical equilibrium}
\label{sec:corr}

Following \cite{Skaugen2018b,Skaugen2018}, we aim at computing the density field $\psi$ so that it remains in elastic equilibrium at all times in the presence of mobile dislocations. Besides solving Eq.~\eqref{psitime} this requires computing an additional smooth distortion $\mathbf{u}^\delta$ to fulfill mechanical equilibrium, i.e., $\nabla \cdot \boldsymbol{\sigma}=0$ with $\boldsymbol{\sigma}$ the total stress field. In the PFC approach, $\psi(\mathbf{r},t)$ is then replaced by $\psi(\mathbf{r}-\mathbf{u}^\delta,t)$ \citep{Skaugen2018b}. In the APFC approach we only have access to the amplitudes by solving Eq.~\eqref{eq:amplitudetime} from which $\psi$ can be reconstructed. For small displacement $\mathbf{u}^\delta$ we can write 
\begin{equation}
\begin{split}
\psi'(\mathbf{r}-\mathbf{u}^\delta,t)&\stackrel{\mathbf{u}^\delta \rightarrow 0}{=} 
\psi(\mathbf{r},t)- [\nabla\psi(\mathbf{r},t)]^{\rm T} \cdot \mathbf{u}^\delta + \mathcal{E}(||\mathbf{u}^\delta||^2)\\
& = \sum_\mathbf{q} \left\{ A_\mathbf{q}(\mathbf{r},t) - [\nabla A_\mathbf{q}(\mathbf{r},t)]^{\rm T} \cdot \mathbf{u}^\delta - i \mathbf{q} \cdot \mathbf{u}^\delta A_\mathbf{q}(\mathbf{r},t)  \right\}e^{i\mathbf{q}\cdot \mathbf{r}} + \mathcal{E}(||\mathbf{u}^\delta||^2) \\
& = \sum_\mathbf{q} A_\mathbf{q}'(\mathbf{r},\mathbf{u}^\delta,t) e^{i\mathbf{q}\cdot \mathbf{r}} +\mathcal{E}(||\mathbf{u}^\delta||^2),
\end{split}
\end{equation}
with 
\begin{equation}
A_\mathbf{q}'(\mathbf{r},\mathbf{u}^\delta,t) = (1-i\mathbf{q} \cdot \mathbf{u}^\delta ) A_\mathbf{q}(\mathbf{r},t) - [\nabla A_\mathbf{q}(\mathbf{r},t)]^{\rm T} \cdot \mathbf{u}^\delta,
\label{eq:ampcorr1}
\end{equation}
which are the amplitudes modified by the deformation $\mathbf{u}^\delta$ and need to be computed. Under the assumption of slowly varying amplitudes, as required by the APFC approach, this quantity is simply given by \citep{Spatschek2010}
\begin{equation}
A_\mathbf{q}'(\mathbf{r},\mathbf{u}^\delta,t) = A_\mathbf{q} ( \mathbf{r},t) e^{-i \mathbf{q} \cdot   \mathbf{u}^\delta},
\label{eq:ampcorr2}
\end{equation}
as
\begin{equation}
A_\mathbf{q} ( \mathbf{r},t) e^{-i \mathbf{q} \cdot   \mathbf{u}^\delta} = A_\mathbf{q} ( \mathbf{r},t) [\cos( \mathbf{q} \cdot   \mathbf{u}^{\delta}) - i \sin(\mathbf{q} \cdot   \mathbf{u}^{\delta})] \stackrel{\mathbf{u}^\delta \rightarrow 0}{\approx} (1-i\mathbf{q} \cdot \mathbf{u}^\delta ) A_\mathbf{q}(\mathbf{r},t),
\label{eq:ampcorr3}
\end{equation}
where the last expression corresponds to Eq.~\eqref{eq:ampcorr1} for negligible gradients of amplitudes.  
In the following we will use Eq.~\eqref{eq:ampcorr2}. A comparison with approximations in Eq.~\eqref{eq:ampcorr1} and \eqref{eq:ampcorr3} will be given in Sect.~\ref{sec:grain-shrinkage}.

As shown in \citep{Skaugen2018b}, $\mathbf{u}^\delta$ can be determined through a Helmholtz decomposition into curl- and divergence-free parts\footnote{We use the notation convention on implicit summations over repeated indices.}, 
\begin{equation}
    u_i^\delta=\partial_i \varphi +\epsilon_{ij} \partial_j \alpha,
    \label{eq:disp}
\end{equation}
with $\varphi$ and $\alpha$ to be determined. $\varphi$ can be computed from a smooth strain $\e_{ij}^\delta$ as
\begin{equation}
    \nabla^2 \varphi = \rm{Tr}(\boldsymbol{\e}^\delta).
    \label{eq:potV}
\end{equation}
The same holds for $\alpha$ and read 
\begin{equation}
    \nabla^4 \alpha = -2\epsilon_{ij}\partial_{ik} \e_{jk}^\delta.
    \label{eq:potA}
\end{equation}
The strain field $\e_{ij}^\delta$ is compatible, and the corresponding stress can be computed from the difference between the total stress, $\boldsymbol{\sigma}$, and a stress computed from the amplitude functions, $\s_{ij}^\psi$, as \citep{Skaugen2018}
\begin{equation}
\s^{\delta}_{ij}=\s_{ij}-\s^\psi_{ij}=\epsilon_{ik}\epsilon_{jl}\partial_{kl}\chi-\s_{ij}^\psi,
\label{eq:sdelta}
\end{equation}
where $\chi$ is the Airy stress function. Exploiting the formulation reported in \cite{Skaugen2018}, the stress $\s_{ij}^\psi$ can be obtained by
\begin{equation} 
    \sigma_{ij}^\psi = \langle \sigma_{ij} \rangle = \langle (\partial_i \mathcal{L}\psi) \partial_j\psi - \mathcal{L}\psi \partial_{ij}\psi \rangle,
    \label{eq:sigma_1}
\end{equation}
where $\mathcal{L}\equiv 1+\nabla^2$ and $\langle \cdots \rangle$ the average over the unit cell, here necessary to compute $\sigma_{ij}^\psi$ in terms of $A_j$. Using Eq.~\eqref{eq:psi} 
and by integrating over the unit cell we 
obtain
\begin{equation}
    \sigma_{ij}^\psi=\sum_\mathbf{q}
    \left\{ 
    [(\partial_i + i q_i)(\nabla^2 + 2i\mathbf{q}\cdot \nabla)A_\mathbf{q}] [(\partial_j - i q_j)A_\mathbf{-q}] 
    -  [(\nabla^2 + 2i\mathbf{q}\cdot \nabla) A_\mathbf{q}] [(\partial_i - i q_i)(\partial_j - i q_j)A_\mathbf{-q}] 
    \right\} .
    \label{eq:sigmapsi1}
\end{equation}
In Eq.~\eqref{eq:sdelta} $\chi$ is given by
\begin{equation}
    (1-\kappa) \nabla^4 \chi =2\mu \epsilon_{ij}\partial_{i}B_j(\mathbf{r})= (\epsilon_{ik}\epsilon_{jl} \partial_{ij}\sigma_{kl}^\psi-\kappa \nabla^2 \sigma_{kk}^\psi),
\label{eq:airy}
\end{equation}
with $\mathbf{B}(\mathbf{r})$ the Burgers vector density, and $\kappa = \lambda/(2(\lambda + \mu))$, with $\lambda$ and $\mu$ the two Lam\'e coefficients. Once $\s_{ij}^\delta$ is computed, the smooth strain to be used in Eqs.~\eqref{eq:potV} and \eqref{eq:potA} is obtained by
\begin{equation}
    \e_{ij}^\delta=\frac{1}{2\mu}(\s_{ij}^\delta - \nu \delta_{ij}\mathrm{Tr}(\boldsymbol{\sigma}^\delta)),
    \label{eq:edelta}
\end{equation}
and, in turn, the smooth deformation $\mathbf{u}^\delta$ is determined from Eqs.~\eqref{eq:disp}--\eqref{eq:potA}. Once $\mathbf{u}^\delta$ is known, Eq. \eqref{eq:ampcorr2} can be used to update the amplitudes. 

We note that, in our formulation, although the stress $\sigma_{ij}$ is required to satisfy mechanical equilibrium at each time step, the configuration of the system changes as dislocations move in time. Dislocation motion and net plastic flow follow from the dissipative evolution of the phase field. The separation of time scales between dissipative plastic motion and elastic relaxation is a basic assumption in most modeling approaches, such as, for instance, discrete dislocation dynamics \citep{Kubin92,Devincre1992}. The strain rate dependence of plastic deformation can then be still maintained in the modeling with instantaneous elastic equilibrium, as demonstrated in e.g., recent PFC study of mechanical deformation of graphene grain boundaries \citep{Zhou19b}.

\section{Finite element implementation}
\label{sec:apfc_fem}

The implementation of the system of partial differential equations (PDEs) reported in Sect.~\ref{sec:corr} builds on the discretization of the standard APFC model described in \cite{SalvalaglioPRE2017,Praetorius2019} and extends it to maintain mechanical equilibrium. It is based on the adaptive FEM toolbox AMDiS \citep{Vey2007,Witkowski2015}. The governing equations are solved as systems of second-order PDEs with semi-implicit integration schemes. For the sake of clarity, these integration schemes are reported in the following in matrix form $\mathbf{L} \cdot \mathbf{x}=\mathbf{R}$, with $\mathbf{x}$ the vector of unknowns. We define an auxiliary complex field $B_j= (\nabla^2 + 2i\qv \cdot \nabla)A_j=\mathcal{G}_j A_j$ and explicitly consider the real and imaginary part of $A_j$ and $B_j$ such as $A_j=\ra + i \ia$ and $B_j=\rx + i \ix$. The following system of equations is then used for numerically integrating the evolution equation \eqref{eq:amplitudetime} for the amplitude $A_j$
\begin{equation}
\mathbf{L}=
\begin{bmatrix}
-\nabla^2 & \mathcal{P} &1 &0 \\[0.75em]
-\mathcal{P} &-\nabla^2 &0 &1 \\[0.75em]
G_1(\{A_i^{(n)}\}) & 0 & K \nabla^2 & -K \mathcal{P} \\[0.75em] 
0 & G_2(\{A_i^{(n)}\}) & K \mathcal{P} & K \nabla^2  
\end{bmatrix} \qquad
 \mathbf{x}=
\begin{bmatrix}
\ra^{(n+1)} \\[0.75em]
\ia^{(n+1)} \\[0.75em] 
\rx^{(n+1)}  \\[0.75em]
\ix^{(n+1)}
\end{bmatrix} \qquad
\mathbf{R}= 
\begin{bmatrix}
0\\[0.75em]
0\\[0.75em] 
H_1(\{A_i^{(n)}\})\\[0.75em]
H_2(\{A_i^{(n)}\}) \\
\end{bmatrix}
\label{eq:integrationschemexR}
\end{equation}
where $n$ is the time step index, $\tau_n>0$ is the time step size at step $n$, $\mathcal{P}=2\mathbf{q}_j\cdot \nabla$, $K=|\mathbf{q}_j|^2B_0^x$, and
\begin{equation}
\begin{split}
G_1(\{A_i\})=&\dfrac{1}{\tau_n}+|\mathbf{q}_j|^2\Delta B+3v|\mathbf{q}_j|^2\left(\Phi+\ra^2-\ia^2\right),\\
G_2(\{A_i\})=&\dfrac{1}{\tau_n}+|\mathbf{q}_j|^2\Delta B+3v|\mathbf{q}_j|^2\left(\Phi+\ia^2-\ra^2\right), \\
H_1(\{A_i\})=&\left[ \dfrac{1}{\tau_n}+6|\mathbf{q}_j|^2v \ra^2 \right]\ra -|\mathbf{q}_j|^2\text{Re}\left(\frac{\delta f^{\rm tri}}{\delta A_j^*}\right), \\
H_2(\{A_i\})=&\left[\dfrac{1}{\tau_n} +6|\mathbf{q}_j|^2v \ia^2 \right]\ia -|\mathbf{q}_j|^2\text{Im}\left(\frac{\delta f^{\rm tri}}{\delta A_j^*}\right).
\end{split}
\label{eq:system}
\end{equation}
$\{A_i\}$ refers to the entire set of amplitudes $A_i$ with $i=1,...,N$ as they enter $\Phi$ and $f^{\rm tri}$.
Exploiting the definition of $B_j$ reported above, the components of the stress $\boldsymbol{\sigma}^\psi$ can be written as
\begin{equation}
\begin{split}
    \sigma_{lm}^\psi=
    &
    \left\{
    [(\partial_l + i q_l)B_j ] [(\partial_m - i q_m)A_j^*] 
    -  B_j [(\partial_l - i q_l)(\partial_m - i q_m)A_j^*] 
     + {\rm c.c.} 
     \right\}
    \\
    %
    =2&
     \left[
        \partial_l \rx \partial_m \ra 
        +\partial_l \ix \partial_m \ia
        +2 \qi \rx \partial_m \ia 
        -2 \qi \ix \partial_m \ra
        - \qj \ia \partial_l \rx
        + \qj \ra \partial_l \ix 
        \right. 
        \\
        &
        \left.
        +2 \qi\qj \ra \rx 
        + 2 \qi\qj \ia \ix
        -\rx\partial_{lm}\ra
        -\ix\partial_{lm}\ia
        + \qj \rx \partial_l \ia 
        - \qj \ix \partial_l \ra \right]
\end{split}
\label{eq:numdevs}
\end{equation}
where the real and imaginary parts of $A_j$ and $B_j$ have been considered.
Note that Eq. \eqref{eq:numdevs} contains the variables numerically computed by the system in Eq.~\eqref{eq:integrationschemexR}, along with their first and second derivatives. Therefore, the stress can be determined directly from the results of the evolution equations at every time step, without explicitly computing the third-order equation \eqref{eq:sigmapsi1}. The semi-implicit integration scheme used to solve the fourth-order PDE in Eq.~\eqref{eq:airy} for $\chi$ reads 
\begin{equation}
\mathbf{L}=
\begin{bmatrix}
\nabla^2 & -1  \\[0.65em]
0 & (1-\kappa) \nabla^2
\end{bmatrix}, \quad
 \mathbf{x}=
\begin{bmatrix}
\chi \\[0.75em]
\eta
\end{bmatrix}, \quad
\mathbf{R}= 
\begin{bmatrix}
0 \\[0.75em]
\mathcal{Q}-\kappa \nabla^2 \s_{ii}^\psi
\end{bmatrix},
\label{eq:integrationscheme1}
\end{equation}
where 
\begin{equation}
    \mathcal{Q}=\epsilon_{ik}\epsilon_{jl} \partial_{ij}\sigma_{kl}^\psi \stackrel{2D}{=}\partial_{xx}\sigma_{yy}^\psi+\partial_{yy}\sigma_{xx}^\psi-\partial_{xy}\sigma_{yx}^\psi-\partial_{yx}\sigma_{xy}^\psi,
    \label{eq:syschi}
\end{equation}
and $\eta$ an auxiliary variable such as $\eta=\nabla^2 \chi$. The FEM calculation for $\chi$ is also exploited to compute the second derivatives of this field to be used in Eq.~\eqref{eq:sdelta} (additional auxiliary variables $\xi_{ij}$ such as $\xi_{ij}=\partial_{ij}\chi$ may be added to the system Eq.~\eqref{eq:syschi}). In turn, $\e_{ij}^\delta$ can be computed straightforwardly from Eqs.~\eqref{eq:edelta}. $\varphi$ and $\alpha$ can then be determined from $\e_{ij}^\delta$. The semi-implicit integration scheme used to solve the fourth-order PDE in Eq.~\eqref{eq:potA} yielding $\alpha$ reads 
\begin{equation}
\mathbf{L}=
\begin{bmatrix}
\nabla^2 & -1  \\[0.65em]
0 & \nabla^2

\end{bmatrix}, \quad
 \mathbf{x}=
\begin{bmatrix}
\alpha \\[0.75em]
\zeta
\end{bmatrix}, \quad
\mathbf{R}= 
\begin{bmatrix}
0 \\[0.75em]
\mathcal{S}
\end{bmatrix},
\label{eq:integrationscheme2}
\end{equation}
where 
\begin{equation}
    \mathcal{S}=-2\epsilon_{ij}\partial_{ik}\e_{jk}^{\delta}
    \stackrel{2D}{=}-2\partial_{xx}\e_{yx}^{\delta}-2\partial_{xy}\e_{yy}^{\delta}+2\partial_{yx}\e_{xx}^{\delta}+2\partial_{yy}\e_{xy}^{\delta},
\end{equation}
and $\zeta$ an auxiliary variable such as $\zeta=\nabla^2 \alpha$. The Poisson equation for $\varphi$ is straightforwardly implemented as a single second-order equation \eqref{eq:potV}. With values of $\alpha$ and $\varphi$, $\mathbf{u}^\delta$ can then be computed from Eq.~\eqref{eq:disp} and the amplitudes updated as reported in Sect.~\ref{sec:corr}, namely by Eq.~\eqref{eq:ampcorr2}. Vanishing potentials at the boundaries are ensured by Dirichlet boundary conditions, while both periodic boundary conditions and no-flux Neumann boundary conditions can be used for amplitudes equations as the systems considered here have constant values for $A_j$ at the boundaries. In the FEM framework, quadratic basis function are used to allow for explicit evaluation of first- and second-order derivatives when needed. Numerical solutions reported below are performed with time steps in the range of $\tau_n=0.1,\ldots,1$, by using spatial mesh adaptivity (see \citet{SalvalaglioPRE2017,Praetorius2019} for more information). 

More general boundary conditions can be implemented in this APFC approach. Traction boundary conditions have been introduced earlier in the PFC model for the study of mechanical deformation through an extra free energy penalty term proportional to $(\psi-\psi_{\rm trac})^2$, where $\psi_{\rm trac}$ is the fixed PFC density imposed on several atomic layers at the boundaries \citep{re:berry06,re:stefanovic09,re:adland13}. This extra term can be used to, e.g., apply a constant shear \citep{re:berry06,re:adland13} or uniaxial tensile load \citep{re:stefanovic09}. It would be straightforward to adopt a similar procedure on the APFC by imposing an energy penalty in terms of complex amplitudes. On the other hand, general scenarios of traction boundary conditions are straightforward to implement in this approach as we work directly with the stress field. Specific values of tractions $\mathbf{T} = \boldsymbol{\sigma} \cdot \hat{\mathbf n}$ (where $\hat{\mathbf n}$ is the boundary surface normal) can be imposed as boundary conditions on the Airy stress function $\chi$ governed by Eq. (\ref{eq:airy}), given that $T_i = \sigma_{ij} \hat{n}_j = (\epsilon_{ik}\epsilon_{jl}\partial_{kl}\chi) \hat{n}_j$.

\section{Numerical results}
\label{sec:results}

\subsection{Single Dislocation in a finite crystal}
\label{sec:single}

We first consider a configuration that includes a single dislocation. The amplitudes are initialized by considering the displacement field given by an edge dislocation with Burgers vector parallel to the $x$-direction, i.e., $\mathbf{b}=(b_x,0)$ with $b_x=a=4\pi/\sqrt{3}$ and $a$ the lattice spacing. Amplitudes are set by \citep{Spatschek2010,SalvalaglioPRE2017}
\begin{equation}
A_j=\phi \exp \left(i \mathbf{q}_j \cdot \mathbf{u} \right),
\label{eq:etafromu}
\end{equation}
where $\phi$ is the real value of amplitudes for a relaxed and unrotated crystal that can be determined through free energy minimization by assuming constant and real amplitudes \citep{ElderPRE2010,SalvalaglioPRE2017}, and $\mathbf{u}$ the displacement field. The components of $\mathbf{u}$ for an edge dislocation are given by \citep{anderson2017}
\begin{equation}
    \begin{split}
        u_x&= \frac{b}{2\pi} \left[ \arctan{\left(\frac{y}{x}\right)} +\frac{xy}{2(1-\nu)(x^2+y^2)} \right],\\
        u_y&= -\frac{b}{2\pi} \left[ \frac{(1-2\nu)}{4(1-\nu)}\log{\left( x^2+y^2 \right)}+\frac{x^2-y^2}{4 (1-\nu) (x^2+y^2)} \right],\\
    \end{split}
    \label{eq:udislo}
\end{equation}
with $\nu$ the Poisson's ratio. Since the displacement field components in Eq.~\eqref{eq:udislo} are singular at the origin, corresponding to the nominal position of the dislocation core \citep{anderson2017}, a local smoothing of the initial condition is introduced. The elastic energy of a single dislocation in bulk, namely in an ideally infinite crystal, is not finite in two dimensions. Therefore, we embed the amplitudes obtained by Eqs.~\eqref{eq:etafromu} and \eqref{eq:udislo} in a circular grain with radius $R$ surrounded by a disordered (liquid) phase. In practice, for $r=\sqrt{x^2+y^2}>R$ both the real and the imaginary parts of the amplitudes are set to zero. The parameters of the (A)PFC free energy are set as $B_x=0.98$, $v=1/3$, $t=1/2$ and, to allow for the coexistence of the solid and liquid phases, $\Delta B_0 \lesssim 8t^2/(135v)$. Having a single dislocation in the system shifts the free energy of the solid to a slightly higher value; thus $\Delta B_0$ should be slightly smaller than that of the ideal, dislocation-free case. Still, we have verified that a negligible growth velocity of the grain is observed for $\Delta B_0=0.042$ over the timescale of interest (i.e., during the relaxation of the system).

\begin{figure}[H]
    \centering
    \includegraphics[width=\textwidth]{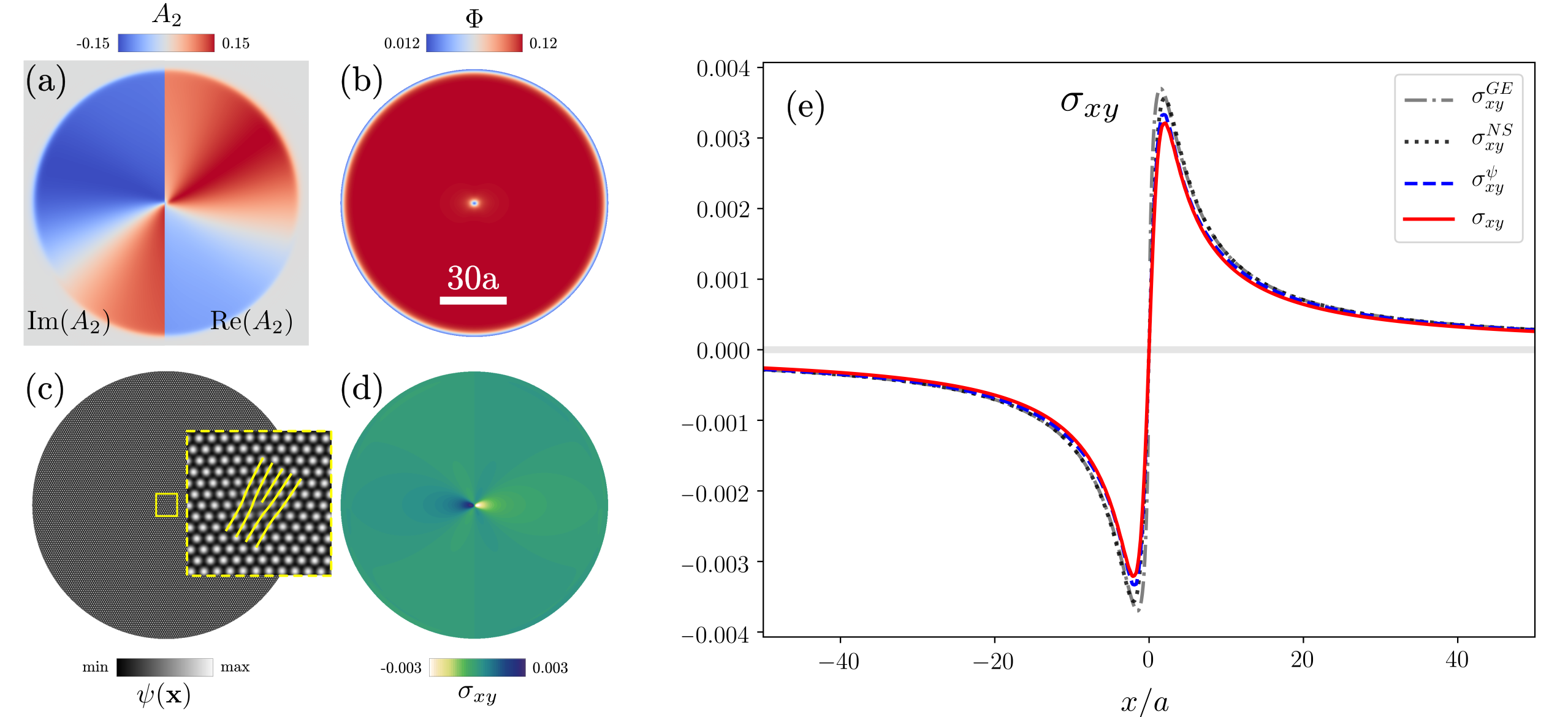}    
    \caption{Computation for a configuration with one edge dislocation in a finite crystal, with $\mathbf{b}=(a,0)$. (a) Real (right) and imaginary (left) parts of the amplitude $A_2$. (b) $\Phi=2\sum_j^N|A_j|^2$. (c) Reconstructed $\psi(\mathbf{x})$ from Eq.~\eqref{eq:psi}, with an inset showing a magnification of the small region around the defect. (d) $\sigma_{xy}$. Panels (a)--(d) are obtained with $R=60a$. (e) Comparison between $\sigma_{xy}$, $\sigma_{xy}^\psi$, $\sigma_{xy}^{\rm NS}$ and $\sigma_{xy}^{\rm GE}$ along the horizontal line crossing the defect core, for $R=300a$.}
    \label{fig:figure1}
\end{figure}

The results obtained at $t=100$ are shown in Fig.~\ref{fig:figure1}(a)--(d): $A_2$ (both real and imaginary parts), $\Phi$ (see Eq.~\eqref{eq:energyamplitude}), $\psi(\mathbf{x})$, and $\sigma_{xy}$ for $R=60a$. Figure \ref{fig:figure1}(e) shows the distribution of $\sigma_{xy}^\psi$, from Eq.~\eqref{eq:sigmapsi1}, and $\sigma_{xy}$ from Eq.~\eqref{eq:sdelta} along the horizontal line through the dislocation core. Here we set $R=300a$ to avoid significant influence of the solid-liquid interface, and of the finite size of the crystal on the elastic field of the dislocation. Since the configuration shown is in equilibrium, $\sigma_{xy}^\psi$ and $\sigma_{xy}$ should be identical. Indeed, the relaxation of the initial state recovers the equilibrium elastic field without the amplitude correction. A very small shift of the stress field is obtained at large distance, which can be ascribed to the presence of the solid-liquid interface. In terms of dislocation self-energy, computed as $\int \boldsymbol{\sigma}^d:\boldsymbol{\varepsilon}^d d\mathbf{x}$ with $\boldsymbol{\sigma}^d$ and $\boldsymbol{\varepsilon}^d$ the stress and strain in the system with a single dislocation, a difference of $\sim 1.5\%$ is found when adding $\boldsymbol{\sigma}^\delta$ to $\boldsymbol{\sigma}^\psi$. It is worth mentioning that the presence of free surfaces, or in general interfaces in inhomogeneous media, affects some  features of the elastic field far from the core \citep{Head_1953,Marzegalli2013,anderson2017}. However, a close comparison to this case is beyond the scope of the present investigation. 

\begin{figure}[H]
    \centering
    \includegraphics[width=\textwidth]{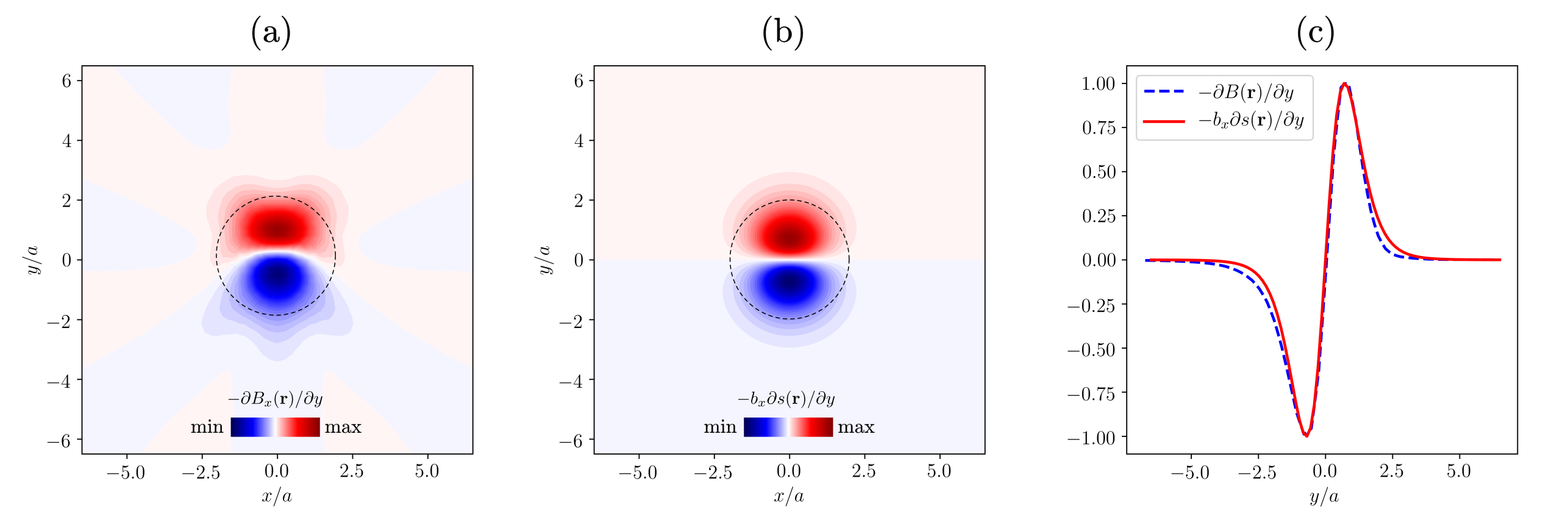}    
    \caption{Analysis of the spreading of $\mathbf{b}$. $\nabla \times \mathbf{B}(\mathbf{r})$ is used to illustrate the spreading of the Burgers vector density given by the phase field stress fields, including results of (a) from APFC simulation of a single dislocation, namely the right-hand side of Eq.~\eqref{eq:airy}, and (b) from Eq.~\eqref{eq:spreadingf} with $c=2b_x$. A circle with radius $c$ is superposed at the origin of plots in panels (a) and (b). (c) Comparison along the vertical line crossing the dislocation core in panels (a) and (b). All the quantities are normalized with respect of the (symmetric) maxima and minima.}
    \label{fig:figure2}
\end{figure}

The stress field given by the APFC model is non-singular at the dislocation core. Our results for the regularized stress field (Fig.~\ref{fig:figure1}(e)) agree with the non-singular theory illustrated in \cite{Cai2006}, where the components of the stress field are argued to be given by,
\begin{equation}
    \begin{split}
        \sigma_{xx}^{\rm NS}&= -\frac{\mu b_x}{2\pi (1-\nu)} \frac{y (3c^2+3x^2+y^2)}{(c^2+x^2+y^2)^2} \\
        \sigma_{yy}^{\rm NS}&= -\frac{\mu b_x}{2\pi (1-\nu)} \frac{y (c^2-x^2+y^2)}{(c^2+x^2+y^2)^2} \\
        \sigma_{xy}^{\rm NS}&= \frac{\mu b_x}{2\pi (1-\nu)} \frac{x (3c^2+x^2-y^2)}{(c^2+x^2+y^2)^2} 
    \end{split}
    \label{eq:sigmaNS}
\end{equation}
with $\mu$ the shear modulus and $c$ a regularization parameter related to the dimension of the dislocation core. Good numerical agreement with our results is found by setting $c=2b_x$. Therefore, the APFC model naturally includes a regularization of the elastic fields at the dislocation core, which deviates from the singular behavior expected from continuum mechanics \citep{anderson2017}, but without requiring any additional parameters. Eq. \eqref{eq:airy} shows how this is accomplished: the right-hand side of this equation corresponds to $\curlB$ (with $z$ labelling the axis perpendicular to the considered 2D system); that is, for the dislocation considered here $\nabla \times (B_x(\mathbf{r}),0) = -\partial B_x(\mathbf{r}) / \partial y ~\hat{z}$. This quantity, for the system of Fig.~\ref{fig:figure1}, is shown in Fig.~\ref{fig:figure2}(a). $-\partial B_x(\mathbf{r}) / \partial y$ is smooth over a finite size region near the core, indicating the effective spreading of the Burgers vector (instead of the isolated singularity $\mathbf{B}(\mathbf{r})=\mathbf{b}\delta(\mathbf{r}-\mathbf{r}_0)$, as expected by continuum mechanics). The spreading of the Burgers vector over a small region around the core is the basic assumption of \citet{Cai2006}. The procedure allows regularised continuous fields at the dislocation core, while still matching the prediction of standard continuum mechanics away from the core (see Eq.~\eqref{eq:sigmaNS}). 

More specifically, the non singular theory of \citep{Cai2006} introduces a spreading function $s(\mathbf{r})$ given by,
\begin{equation}
    s(\mathbf{r})=\frac{15}{8\pi c^3 (r^2/c^2+1)^{7/2}}.
    \label{eq:spreadingf}
\end{equation}
The convolution of this function with non-singular stress-field components leads to Eqs.~\eqref{eq:sigmaNS}. Therefore, we can assume a spreading of the Burgers vector density such as $\mathbf{B}^{\rm NS}(\mathbf{r})=(B_x^{\rm NS}(\mathbf{r}),0)$ with $B_x^{\rm NS}(\mathbf{r})=b_x s(\mathbf{r})$ for an edge dislocation having $\mathbf{b}=(b_x,0)$. Therefore, $\nabla \times B_x^{\rm NS}(\mathbf{\mathbf{r}})=-b_x\hat{z}\partial s(\mathbf{r}) / \partial y $. It is worth mentioning that in \citet{Cai2006}, the starting point is to assume a regularization of the Burgers vector density by a function $\tilde{s}(\mathbf{r})$, whose convolution with itself gives Eq. \eqref{eq:spreadingf}. The latter enters the deformation fields and allows the removal of the singularity at the core given that $\int r(\mathbf{x}-\mathbf{x'}) s(\mathbf{x'})d^3\mathbf{x'}= r_c$ with $r_c=\sqrt{x^2+y^2+z^2+c^2}$. The distribution of $\nabla \times \mathbf{B}_x^{\rm NS}(\mathbf{\mathbf{r}})$ is shown in Fig.~\ref{fig:figure2}(b) for $c=2b_x$. A circle with radius $2b_x$ is superposed on both the maps in Figs.~\ref{fig:figure2}(a) and (b). In addition, a comparison of the two distributions along the vertical line crossing the defect is shown in Fig.~\ref{fig:figure2}(c). The agreement between them indicates that the assumption of isotropic spreading given in \citet{Cai2006} does capture the main features of the regularization at the core given by the APFC model. However, it should be noted that the APFC description naturally incorporates the lattice symmetry and as such will always include any anisotropies in this (or any other) quantity. 

Another regularization scheme similar to that discussed above has been introduced by assuming a smooth Burgers vector distribution \citep{Lothe1992}, or by elasticity theories including first strain-gradient energy terms \citep{Mindlin1964,Mindlin1968}. In the so-called Helmoltz type gradient elasticity, for an edge dislocation with Burgers vector oriented along the $x$-axis (i.e., $\mathbf{b}=(b_x,0)$), the elastic field in an isotropic medium is given by \citep{Lazar2005,Lazar2017}
\begin{equation}
    \begin{split}
    \sigma_{xx}^{\rm GE}&=-\frac{\mu b_x}{2\pi(1-\nu)}\frac{y}{r^4}\left[(y^2+3x^2)+\frac{4\ell^2}{r^2}(y^2-3x^2)-2y^2\frac{r}{\ell}K_1(r/\ell)-2(y^2-3x^2)K_2(r/\ell)\right], \\
    \sigma_{yy}^{\rm GE}&=-\frac{\mu b_x}{2\pi(1-\nu)}\frac{y}{r^4}\left[(y^2-x^2)-\frac{4\ell^2}{r^2}(y^2-3x^2)-2x^2\frac{r}{\ell}K_1(r/\ell)+2(y^2-3x^2)K_2(r/\ell)\right],\\
    \sigma_{xy}^{\rm GE}&=\frac{\mu b_x}{2\pi(1-\nu)}\frac{x}{r^4}\left[(x^2-y^2)-\frac{4\ell^2}{r^2}(x^2-3y^2)-2y^2\frac{r}{\ell}K_1(r/\ell)+2(x^2-3y^2)K_2(r/\ell)\right],\\
    \end{split}
    \label{eq:sigmaGE}
\end{equation}
with $K_n(r/\ell)$ the modified Bessel function of the second type, and $\ell$ a characteristic internal length parameter of the material. This parameter is usually proportional to the lattice spacing, as has been obtained empirically by comparison with atomistic calculation (see, e.g., \citep{Po2014}). The elastic field obtained in this approach is also shown in Fig.~\ref{fig:figure1} for $\ell=b_x$. With this choice of $\ell$, $\sigma_{\rm xy}^{\rm GE} \approx \sigma_{xy}^{\rm NS}$ within the core region. In turn, both agree with $\sigma_{xy}$ obtained from the APFC model. Within first-gradient elasticity, the spreading of the Burgers vector is given by a function that is singular at the core, $K_0(r/\ell)\approx -\left\{\log[r/(2\ell)]+\gamma\right\}\left\{1+r^2/(4\ell^2)\right\}-r^2/(4\ell^2)+\mathcal{O}(r^4)$ with $\gamma=0.57721...$ the Euler-Mascheroni constant \citep{Lazar2017}. Therefore an analysis analogous to that of Fig.~\ref{fig:figure2} is not possible. Closer comparisons with this theory will be the subject of future research.

These comparisons shed light on the regularization of stress fields in the (A)PFC framework. Other than the agreement described above, we note that small deviations and asymmetries are observed, in particular concerning the highest stress values obtained in the system (see the curves in Fig.~\ref{fig:figure1}(e)). This is, however, expected as nonlinearities are contained in PFC amplitudes \citep{Huter2016} which generally capture features on atomic length scales. They would become relevant for large stresses/strains. It is worth mentioning that regularization of the elastic field at the core of a dislocation is natural from an atomistic point of view, as the distribution is expected to be non-singular with vanishing deformation field at the core (see, e.g., \cite{Bonilla2015}) which intrinsically has a finite size. This was also the main argument that led to the renowned Peirels-Nabarro model \citep{Peierls_1940,Nabarro_1947}, although in this case the resulting elastic fields are discontinuous at the core, even if they do not diverge \citep{Lazar2017}. Therefore they deviate from the continuous description given by the APFC model.

\subsection{Dislocation dipole}
\label{sec:dipole}

In this section a dislocation dipole is considered. The amplitudes are initialized by considering the displacement field of two edge dislocations with Burgers vector $b_x=\pm a$ aligned along the $x$ direction. Eqs.~\eqref{eq:etafromu} and \eqref{eq:udislo} are straightforwardly used by considering $\mathbf{u}(x-x_0,y-y_0)$, i.e. shifting the $x$ and $y$ axis to account for the initial position of each dislocation $\mathbf{p}=(x_0.y_0)$. We focus here in particular on two defects having positions $\mathbf{p}_{1,2}=(0,\pm L)$ with $L \sim 10a$. Parameters of the (A)PFC free energy are set as those in Sect.~\ref{sec:single}. A square computational domain with size $200a$ is used, embedding a grain as in the previous section with radius $140a$.

\begin{figure}[h]
    \centering
    \includegraphics[width=\textwidth]{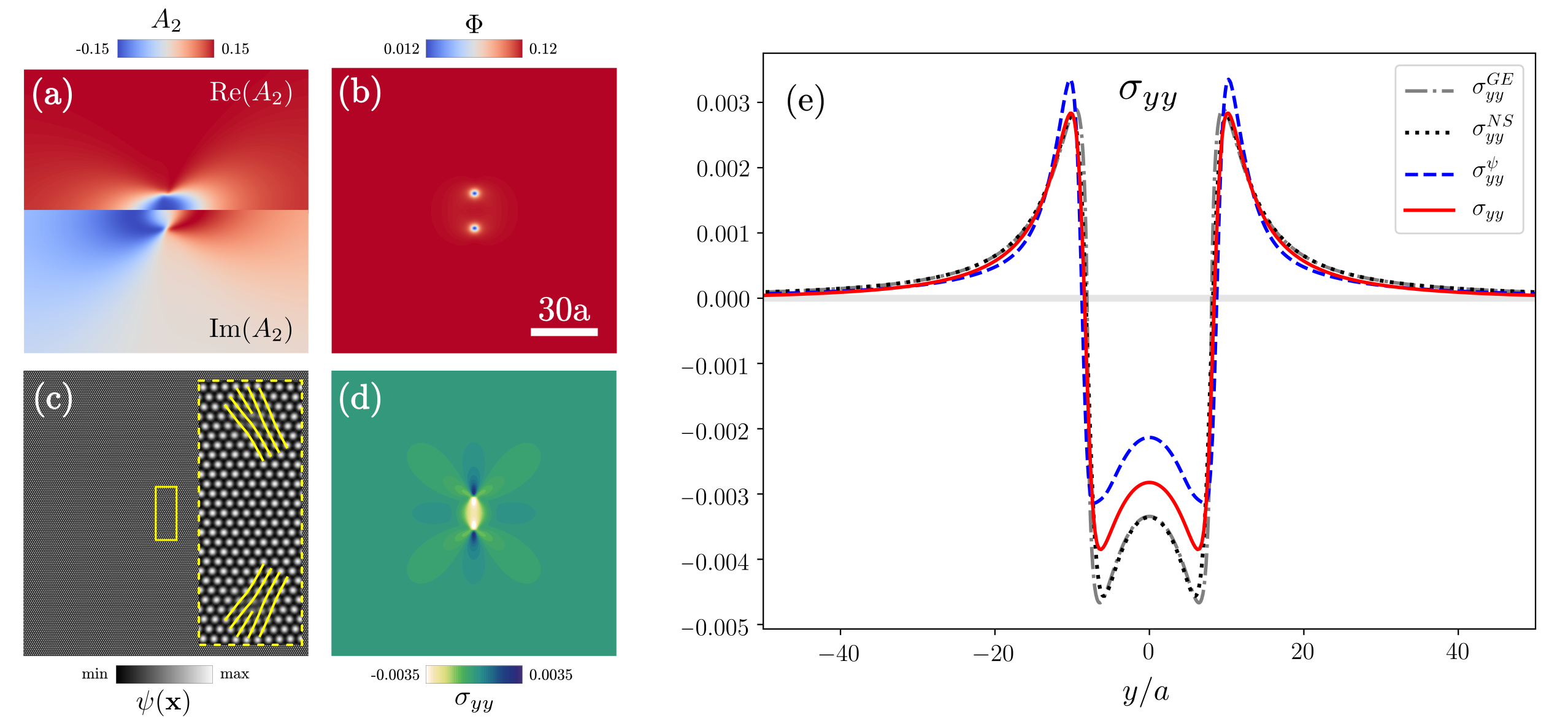}    
    \caption{Simulation of an edge-dislocation dipole with $\mathbf{b}_{1,2}=(\pm a,0)$ and positions $\mathbf{p}_{1,2}=(0,\pm 10a)$. (a) Amplitude $A_2$ illustrated by its real (top) and imaginary (bottom) parts. (b) $\Phi=2\sum_j^N|A_j|^2$. (c) Reconstructed $\psi(\mathbf{x})$ from Eq.~\eqref{eq:psi}, with an inset showing a magnification of the small region around the defect. (d) $\sigma_{yy}$. (e) Comparison between $\sigma_{yy}$, $\sigma_{yy}^\psi$, $\sigma_{yy}^{\rm NS}$, and $\sigma_{yy}^{\rm GE}$ along a vertical line crossing the defects.}
    \label{fig:figure3}
\end{figure}
The results obtained at $t=100$ are shown in Fig.~\ref{fig:figure3}(a)-(d) which illustrate $A_2$, $\Phi$, $\psi(\mathbf{x})$, and $\sigma_{yy}$. The stress fields $\sigma^{\psi}$ and $\sigma$, along with the predictions from non-singular continuum elasticity theories as described in Sect.~\ref{sec:single} are presented in Fig.~\ref{fig:figure3}(e). First, note that the stress $\sigma^{\psi}$ and the total stress $\sigma$ differ from each other, even after a relaxation time from the initial condition which was long enough to obtain a substantial agreement for the case of an isolated dislocation (Sect.~\ref{sec:single}). These APFC results are obtained with parameters corresponding to coexistence of the liquid and solid phases, for which we expect a good description of the elastic relaxation in the standard (A)PFC model, different from deep quenches conditions. The system at $t=100$ is, however, out of equilibrium as the defects will move and finally annihilate. Therefore, as discussed in \citep{Skaugen2018b}, the correction discussed in Sect.~\ref{sec:corr} is needed to maintain mechanical equilibrium. Note that in Fig.~\ref{fig:figure3}(e) the stress fields obtained from continuum elasticity theories, namely Eqs.~\eqref{eq:sigmaNS}--\eqref{eq:sigmaGE}, have been properly shifted to account for the dislocations forming the dislocation dipole, and are expected to match the PFC elastic field far away from the core (e.g., when $|y|>20a$). Closer to the defects some deviations from continuum elasticity are observed. Such deviations are shown in the regimes of large deformation near dislocation cores (even in the single-dislocation case of Fig.~\ref{fig:figure1}). They are more evident in between the individual defects, where the contributions of the two dislocations accumulate. This effect may be ascribed to nonlinearities of elasticity as they are naturally contained in APFC amplitudes and play an important role at high strains \citep{Huter2016}. In general, they become relevant when the distortion of the lattice parameter $a$ compared to the lattice parameter of the reference crystal, $a_{\rm bulk}$, is $a/a_{\rm bulk}<0.95$ or $a/a_{\rm bulk}<1.05$. Here, strains larger than $5\%$ are observed in the region $|y|<20a$.

\subsection{Motion of a dislocation dipole}
\label{sec:dipole-motion}

The configuration of Sect.~\ref{sec:dipole} also allows us to investigate the evolution that satisfies the  constrain of mechanical equilibrium of elastic distortions as discussed in Sect.~\ref{sec:corr}. We consider here a dislocation dipole with dislocations annihilating by pure glide or climb. We choose as initial conditions two dislocations at $\mathbf{p}_{1,2}=(\pm L,0)$ (configuration G, glide), and $\mathbf{p}_{1,2}=(0,\pm L)$ (configuration C, climb), with $L \sim 15a$. The latter system corresponds to the one analyzed in Sect.~\ref{sec:dipole}, shown in Fig.~\ref{fig:figure3}(a)--(d), while the former consists of the same defects but aligned along the y direction.

\begin{figure}[H]
    \centering
    \includegraphics[width=\textwidth]{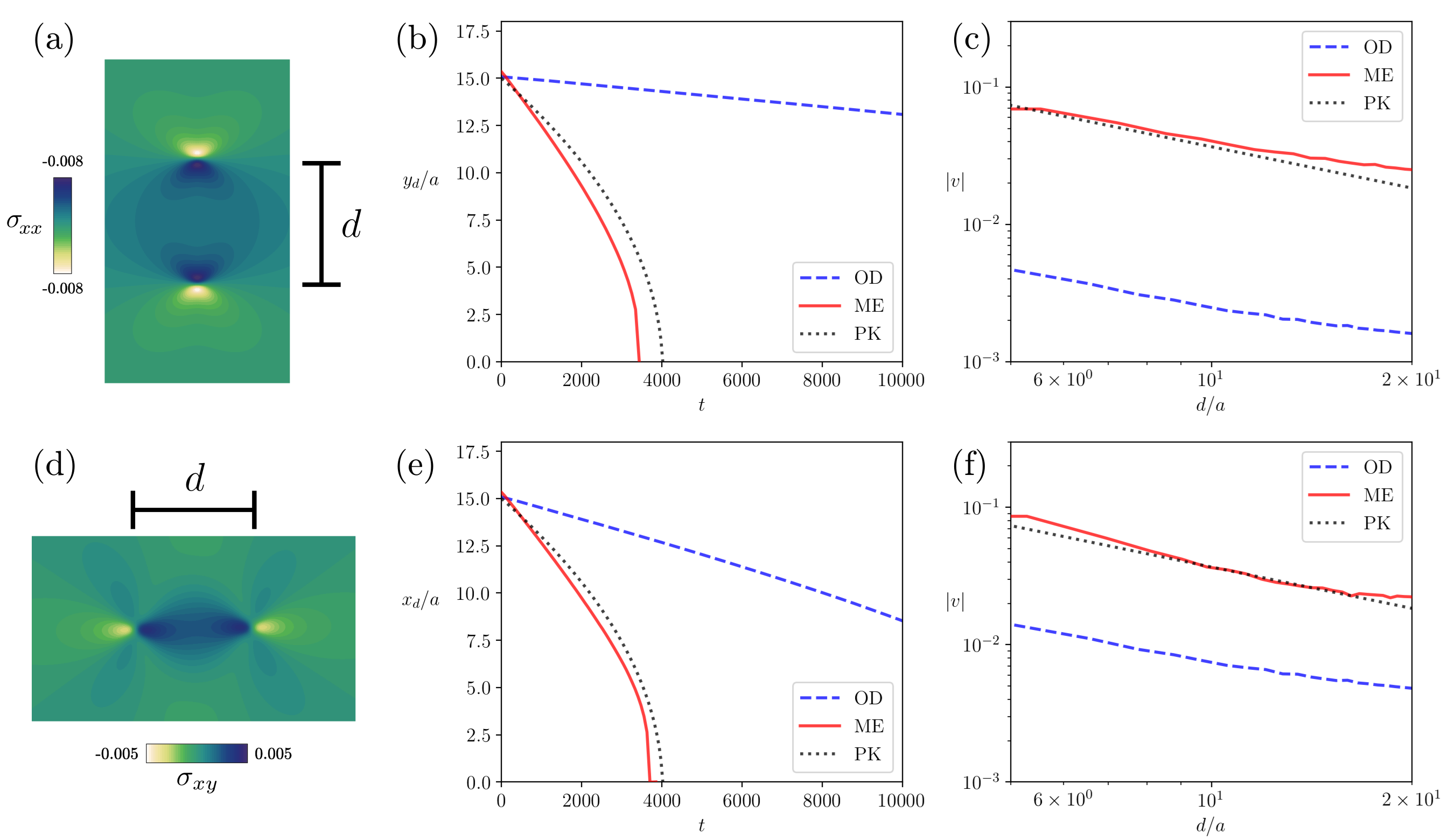}
    \caption{Defect annihilation. For configuration C: (a) $\sigma_{xx}$, (b) position ($y_d/a$) over time over time, and (c) velocity. For configuration G: (d) $\sigma_{xy}$, (e) position ($x_d/a$) over time, and (f) velocity. In this and 
    other figures, OD, ME and PK refer to overdamped dynamics (the original APFC model), mechanical equilibrium (the present work) and the Peach-Kohler result (Eq. (\ref{eq:PK})), respectively.}
    \label{fig:figure4}
\end{figure}

Our results are presented in Fig.~\ref{fig:figure4}. Panels Fig.~\ref{fig:figure4}(a) and Fig.~\ref{fig:figure4}(d) show the stress tensor components $\sigma_{xx}$ and $\sigma_{xy}$. Fig.~\ref{fig:figure4}(b) and Fig.~\ref{fig:figure4}(e) show the position over time of the upper defect for configuration C ($y_d$), and of the defect on the right for configuration G ($x_d$), respectively. Panels (c) and (f) show the velocity for the two configurations. Model parameters are the same as in Sect.~\ref{sec:single} and \ref{sec:dipole}, corresponding to solid/liquid coexistence. As a first observation, a faster dynamics is obtained when enforcing mechanical equilibrium in PFC, in agreement with \citep{Skaugen2018,Skaugen2018b}. From a macroscopic point of view, the motion of a dislocation can be described in terms of the Peach-Koehler force \citep{anderson2017}, that is, $f=(\boldsymbol{\sigma}\cdot \mathbf{b})\times \boldsymbol{\xi}$ with $\boldsymbol{\xi}$ the unit vector oriented along the dislocation line, $\mathbf{b}$ the Burgers vector of the dislocation, and $\boldsymbol{\sigma}$ the external elastic field (for a recent review see \citep{LUBARDA2019}). For the configurations considered here, the force acting on a dislocation is $f_{\rm G}=\sigma_{xy}b$ and $f_{\rm C}=\sigma_{xx}b$, with $\sigma_{ij}$ the stress field generated by the other dislocation. Within the (A)PFC framework, the velocity of the dislocations due to the action of the Peach-Kohler force is given by \citep{Skaugen2018}
\begin{equation}
    v_i^{PK}=\frac{1}{4\pi\phi^2}\epsilon_{ij} \sigma_{jk} b_k.
    \label{eq:PK}
\end{equation}
This velocity is shown in Figs.~\ref{fig:figure4}(b) and (d) for configurations C and G, respectively, using equations given in Sect.~\ref{sec:single} to compute $\sigma_{ij}$. Notice that it is in good agreement with $v_x^{\rm G}=v_y^{\rm C}=b^2/(2\pi^2 d)$ \citep{Skaugen2018} (they match for $\nu=1/4$). Note that the purely diffusive dynamics of the APFC model significantly underestimates the magnitude of the velocities. By constraining the stress field so that the evolving configuration remains in mechanical equilibrium, the computed velocity agrees well with the prediction based on the Peach-Kohler force.

We mention that, from a microscopic point of view, the evolution of configurations C and G are expected to be significantly different. Glide is the movement of dislocations along their slip planes, whereas climb is the motion perpendicular to the slip plane. Both are activated processes over different types of barriers. In the former, a layer of atoms slips over the Peierls-Nabarro barrier, whereas the latter requires the absorption (or emission) of vacancies \citep{anderson2017}. The first effect can be captured in the standard PFC model \citep{re:boyer02b,Skaugen2018} which includes atomic scale microscopic features that are not present in the APFC model \citep{Huang13}. Still, at variance with continuum models based solely on mechanics, temperature is included in the APFC approach in a phenomenological fashion as it enters the energy of the ordered and disordered phases. This can be seen, for instance, when changing values of $\Delta B_0$. We recall that $\Delta B_0=8t^2/(135v)$ allows for the coexistence of solid and liquid phases, $\Delta B_0 \gtrless 8t^2/(135v)$ favors the liquid or solid phase, while $\Delta B_0 \ll 8t^2/(135v)$ corresponds to a deep-quench condition. We have verified that increasing $\Delta B_0$ (up to the solid-liquid coexistence condition) speeds up defect motion thus capturing, at least qualitatively, the expected change in defect mobility, although no barriers are explicitly included in the APFC model.

\subsection{Grain shrinkage}
\label{sec:grain-shrinkage}

\begin{figure}[H]
         \centering
         \includegraphics[width=1.0\textwidth]{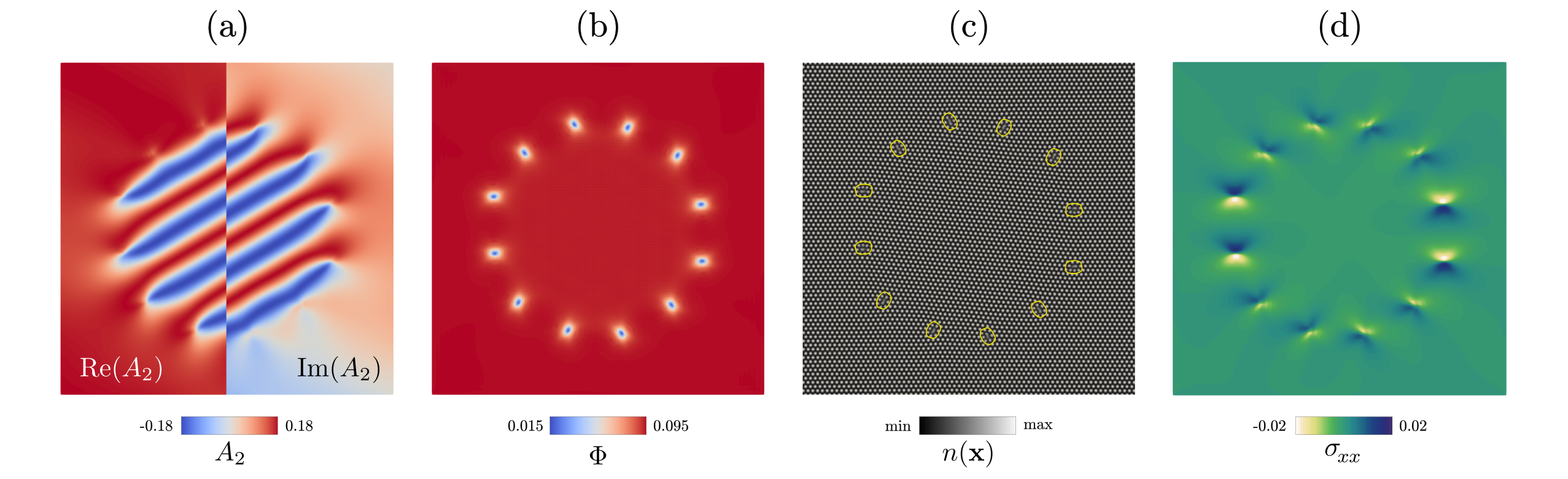}
       \caption{Defects at a circular low-angle grain boundary between a rotated inclusion and an unrotated crystal matrix of triangular symmetry. 
       (a) Amplitude $A_2$ illustrated by its real (left) and imaginary (right) parts. (b) $\Phi=2\sum_j^N|A_j|^2$. (c) Reconstructed $\psi(\mathbf{x})$ from Eq.~\eqref{eq:psi}, with yellow isolines of $\Phi$ highlighting the defects. (d) $\sigma_{xx}$. }
       \label{fig:figure5}
\end{figure}

\begin{figure}[h]
    \centering
    \includegraphics[width=0.6\textwidth]{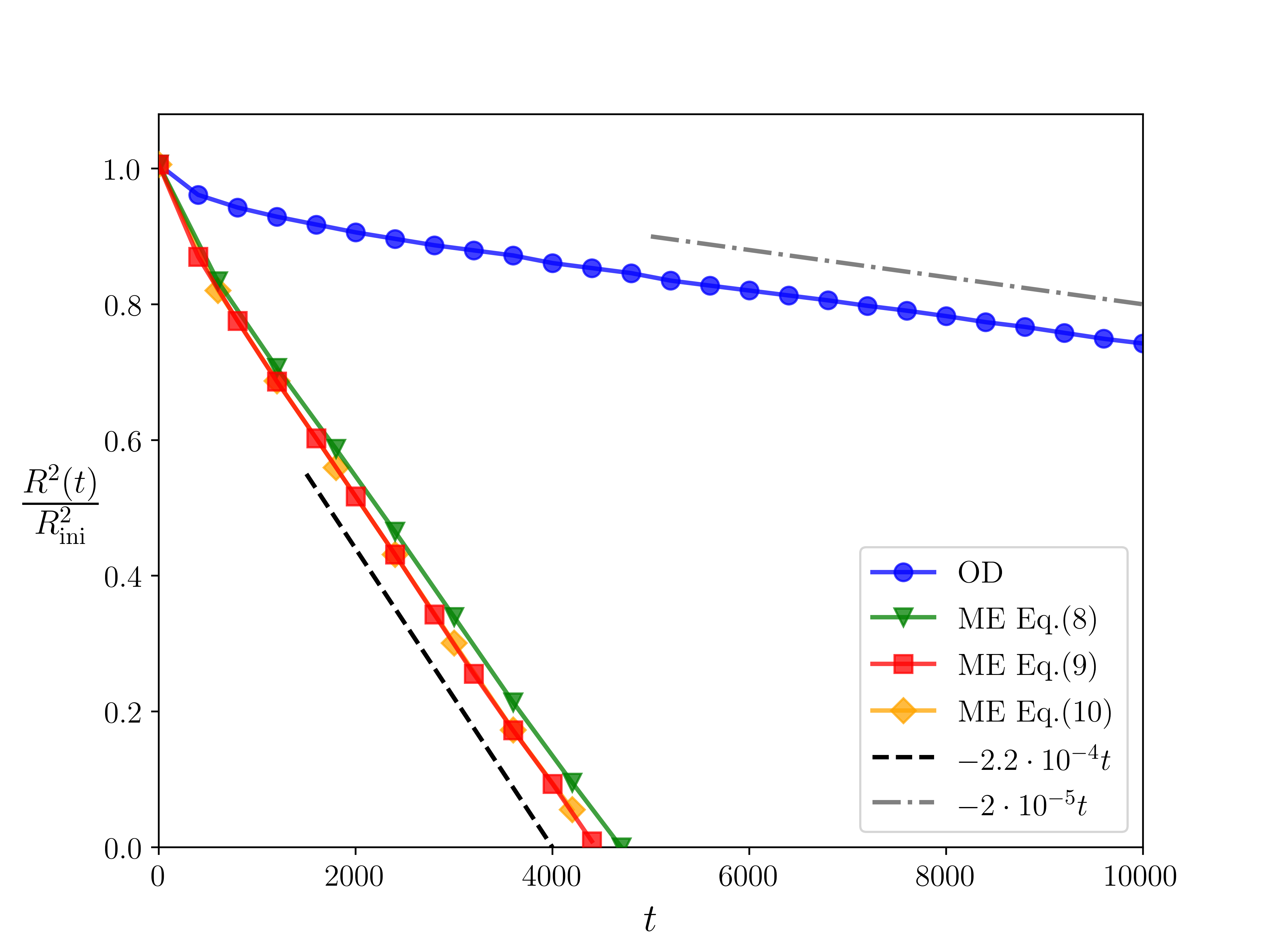}    
    \caption{Time evolution of the normalized area of rotated grain, for the overdamped dynamics (OD) and with the correction to amplitudes for mechanical equilibrium (ME) as in Eqs.~\eqref{eq:ampcorr1}--\eqref{eq:ampcorr3}.}
    \label{fig:figure6}
\end{figure}

We consider here a 2D system with a rotated grain embedded in a crystalline matrix of triangular symmetry. The grain boundary consists of a series of dislocations that move together. Amplitudes are initialized as \citep{SalvalaglioPRE2017}
\begin{equation}
A_j=\phi_j \text{exp}\left(i \delta \mathbf{q}_j (\theta) \cdot \mathbf{x} \right),
\label{eq:etafromtheta}
\end{equation}
with
\begin{equation}
\delta\mathbf{q}_{j}(\theta) =  \left[q^x_{j} (\cos\theta -1) - q^y_{j}
\sin\theta\right]\hat{\mathbf{x}} + \left[q^x_{j} \sin\theta  + q^y_{j}
(\cos\theta-1)\right]\hat{\mathbf{y}} .
\label{eq:krot}
\end{equation}
$\theta = 0$ at a distance $R_0$ from the center of the rotated inclusion/grain, which is tilted by an angle $\theta$ with respect to the surrounding matrix. The spatial distributions of $A_2$, $\Phi$, $\psi(\mathbf{x})$, and $\sigma_{xx}^\psi$, corresponding to $\theta=5^\circ$ and $R_0=25\pi$, are shown in Fig.\ref{fig:figure5}. The parameters of the APFC model are set to be the same as those in the previous sections. The circular grain considered here shrinks over time, which is qualitatively well described by the APFC model \citep{Heinonen2016,Salvalaglio2018,Salvalaglio2019}. Results of the normalized area $R^2(t)/R^2_{\rm ini}$ of the shrinking grain with and without the correction described in Sect.~\ref{sec:corr} are shown in Fig.~\ref{fig:figure6}.

For this configuration, the three expressions of the amplitudes corrected by $\mathbf{u}^\delta$, namely Eqs. \eqref{eq:ampcorr1}, \eqref{eq:ampcorr2} or \eqref{eq:ampcorr3} reported in Sect.~\ref{sec:corr}, are used. In all the cases considered, $R^2(t)$ decreases linearly with time. A significantly faster decrease is observed when mechanical equilibrium is imposed, as has also been observed for the evolution of dislocation dipoles in Sect.~\ref{sec:dipole-motion} and, for this configuration in particular, in \citep{Heinonen2016}. In agreement with our results, the work reported in \cite{Heinonen2016}, in the limit of fast relaxation of elastic excitations, also shows that grain-shrinking dynamics is an order of magnitude faster when accounting for instantaneous mechanical equilibrium. The substantial agreement between three different approaches for correcting the amplitudes (i.e., Eqs.~\eqref{eq:ampcorr1}, \eqref{eq:ampcorr2} and \eqref{eq:ampcorr3}) supports the assumption of small deformations and slowly varying amplitudes. 

\begin{figure}[h]
         \centering
         \includegraphics[width=1.0\textwidth]{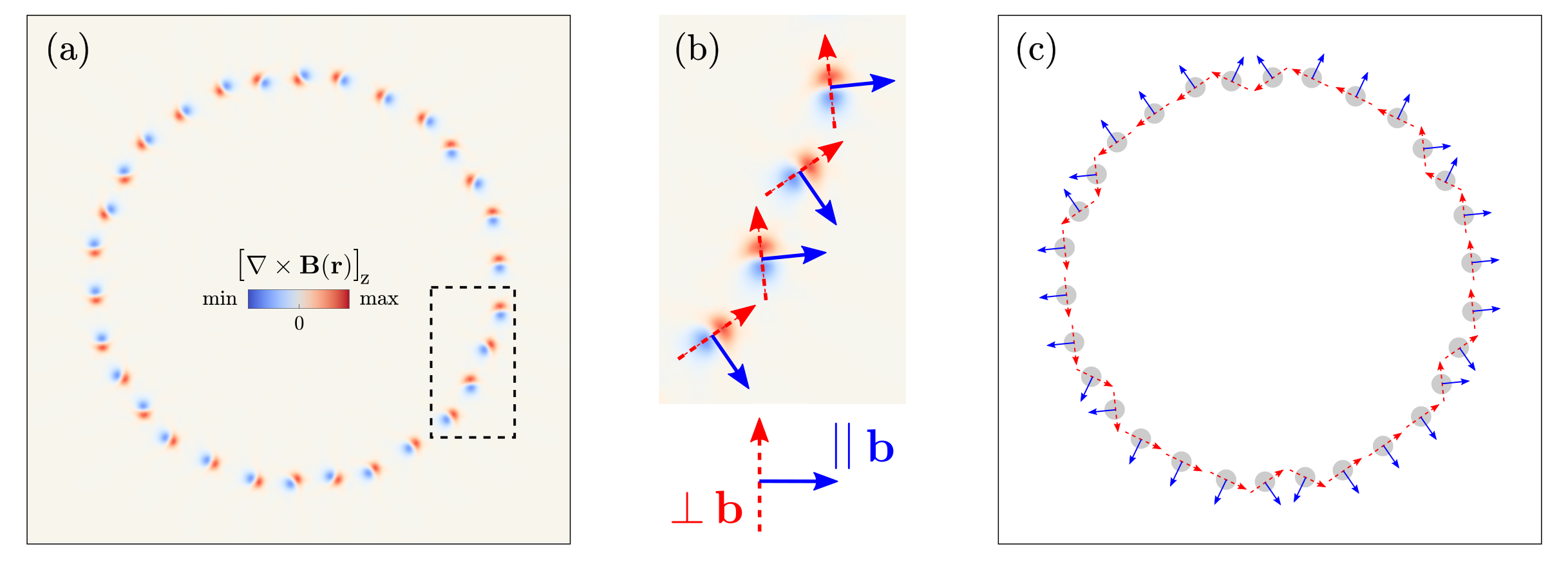}
       \caption{Analysis of the lattice deformation for a rotated inclusion with $\theta=10^\circ$ and $R\sim 20a$. (a) $\curlB$. (b) Detail inside the dashed box of (a), showing the identification of the directions perpendicular (red, dashed arrow) and parallel (blue, solid arrow) to the Burgers vector $\mathbf{b}$. (c) Schematics reporting the scheme of panel (b), where the grey circles (with radius $\sim 2a$) correspond to the positions of individual defects.}
       \label{fig:figure7}
\end{figure}

The analysis of Sect.~\ref{sec:single} concerning how lattice distortion follows from the Burgers vector density $\mathbf{B}(\mathbf{r})$, can be readily applied to this configuration to gain insights about the crystal defects in the rotated inclusion and, more in general, in systems with many defects. The spatial distribution $\curlB$, computed as the right-hand side of Eq.~\eqref{eq:airy} for a rotated inclusion with $\theta=10^\circ$ and $R\sim20a$, is shown in Fig.~\ref{fig:figure7}(a). We obtain a localized distribution centered at defects, as in Fig.~\ref{fig:figure2}(a). By looking at the arrangement of positive and negative lobes of the distribution, six different orientations are obtained, which correspond to multiples of $30^\circ$, consistent with the lattice vectors of the triangular lattice. Indeed, this quantity fully describes the distribution of $\mathbf{b}$ in 2D. Following the arguments of Sect.~\ref{sec:single}, we can identify the orientation of the Burgers vector as being perpendicular to the line connecting the local minimum and maximum of $\curlB$ at defect cores. This is illustrated in Fig.~\ref{fig:figure7}(b). The extension of the local non-zero distributions of $\curlB$ at defects, as well as its maximum and minimum, are then connected to the Burgers vector. Here they are all equivalent as they should yield $|\mathbf{b}|=a$ for symmetry reasons. Fig.~\ref{fig:figure7}(c) gives a schematic illustration of defects with the orientations obtained. Note that the distribution of defects is symmetric, and the sum of all the $\mathbf{b}$'s is zero. It is worth mentioning that for this dynamical system, at variance with Sect.~\ref{sec:single}, the number of defects and the orientation of individual defects are not known a priori. Therefore, this analysis can be used to extract information on the nature of the defects and their evolution towards equilibrium, which is fully contained in the APFC model.

\section{Conclusions}
We have presented a coarse-grained model of lattice distortion and plastic motion as described by the phase-field crystal model in its amplitude expansion formulation. For a system in equilibrium, we have studied the stress field induced by an isolated dislocation, including the regularized stress near the dislocation core. Although the APFC approach cannot provide a detailed description of dislocation cores from an atomistic point of view due to its underlying assumptions \citep{Goldenfeld2005,GoldenfeldJSP2006,Yeon2010}, we have shown that the resulting deformation fields near the core are smooth, and are generally in agreement with other theories based on continuum elasticity \citep{Lazar2005,Cai2006}. Concerning plastic flow, the APFC model has been extended to account for fast elastic relaxation by not only determining the incompatible stress from the phase field, but also introducing a compatible distortion to satisfy elastic equilibrium at all times \citep{Skaugen2018b}. The modified model not only agrees well with predictions from continuum elasticity, but it can also include lattice symmetry, naturally account for the formation and motion of topological defects, and computationally, access large system sizes and long time scales \citep{Salvalaglio2019,Praetorius2019}. Despite the coarse-grained nature of the model, it can provide information about individual defects and defect distributions directly from APFC model variables. In particular, we have shown how to compute the Burgers vector density and its motion from the model amplitudes.

Future work will be devoted to deepen the connections with theories based on continuum mechanics, to extend the results to other lattice symmetries, and to the investigation of three-dimensional systems. The possibility to account for changes in the local average density of the crystal (which is assumed to be constant here) will also be explored. It will allow the extension to the study of binary systems \citep{ElderPRE2010}, and to include a better description of elastic constants \citep{Wang2018,Ainsworth2019}.

\section*{Acknowledgments}
We acknowledge M. Lazar and B. Svendsen for fruitful discussions. A.V. acknowledges support from the German Research Foundation under Grant No. Vo899/20 within SPP 1959. K.R.E. acknowledges financial support from the National Science Foundation (NSF) under Grant No. DMR-1506634, Z.-F.H. acknowledges support from NSF under Grant No. DMR-1609625, and J.V. acknowledges support from NSF under Grant No. DMR-1838977. L.A. acknowledges support from Research Council of Norway through CoE funding scheme, Project No. 262644. We also gratefully acknowledge the computing time granted by the John von Neumann Institute for Computing (NIC) and provided on the supercomputer JURECA at J\"ulich Supercomputing Centre (JSC), within the Project No. HDR06, and by the Information Services and High Performance Computing (ZIH) at TU Dresden.



\end{document}